\documentclass[sigconf]{acmart}

\AtBeginDocument{%
  \providecommand\BibTeX{{%
    \normalfont B\kern-0.5em{\scshape i\kern-0.25em b}\kern-0.8em\TeX}}}

\setcopyright{acmcopyright}
\copyrightyear{2024}
\acmYear{2014}
\acmDOI{10.1145/3649217.3653575}

\acmConference[ITiCSE 2024]{July 8–10, 2024}{Milan, Italy}
\acmISBN{979-8-4007-0600-4/24/07}

\usepackage{todonotes}
\usepackage{outlines}
\usepackage{lipsum}
\usepackage{subcaption}
\usepackage{multirow}

\begin{document}

\title{Student Perspectives on Generative Artificial Intelligence in Computing Education}

\author{C. Estelle Smith}
\orcid{0000-0002-4981-7105}
\affiliation{%
  \institution{Colorado School of Mines, Department of Computer Science}
  \city{Golden}
  \state{CO}
  \postcode{80401}
  \country{USA}}
\email{estellesmith@mines.edu}

\author{Kylee Shiekh}
\orcid{0000-0003-0354-0494}
\affiliation{%
  \institution{Colorado School of Mines, Department of Physics}
  \city{Golden}
  \state{CO}
  \postcode{80401}
  \country{USA}}
\email{knshiekh@mines.edu}

\author{Hayden Cooreman}
\orcid{0009-0004-6082-4455}
\affiliation{%
  \institution{Colorado School of Mines, Department of Computer Science}
  \city{Golden}
  \state{CO}
  \postcode{80401}
  \country{USA}}
\email{hcooreman@mines.edu}

\author{Sharfi Rahman}
\orcid{0009-0007-2212-0774}
\affiliation{%
  \institution{Colorado School of Mines, Department of Computer Science}
  \city{Golden}
  \state{CO}
  \postcode{80401}
  \country{USA}}
\email{sharfirahman@mines.edu}

\author{Yifei Zhu}
\orcid{0000-0001-7802-7869}
\affiliation{%
  \institution{Colorado School of Mines, Department of Computer Science}
  \city{Golden}
  \state{CO}
  \postcode{80401}
  \country{USA}}
\email{zhu1@mines.edu}

\author{Md Kamrul Siam}
\orcid{0000-0001-9986-210X}
\affiliation{%
  \institution{Colorado School of Mines, Department of Computer Science}
  \city{Golden}
  \state{CO}
  \postcode{80401}
  \country{USA}}
\email{siam@mines.edu}

\author{Michael Ivanitskiy}
\orcid{0000-0002-4213-4993}
\affiliation{%
  \institution{Colorado School of Mines, \\
  Applied Mathematics \& Statistics}
  \city{Golden}
  \state{CO}
  \postcode{80401}
  \country{USA}}
\email{mivanits@mines.edu}

\author{Ahmed M. Ahmed}
\orcid{0009-0007-4462-8173}
\affiliation{%
  \institution{Colorado School of Mines, Department of Geophysics}
  \city{Golden}
  \state{CO}
  \postcode{80401}
  \country{USA}}
\email{ahmedmohamedahmed@mines.edu}

\author{Michael Hallinan}
\orcid{0009-0000-0212-4873}
\affiliation{%
  \institution{Colorado School of Mines, \\ 
  Chemical \& Biological Engineering}
  \city{Golden}
  \state{CO}
  \postcode{80401}
  \country{USA}}
\email{mhallinan@mines.edu}

\author{Alexander Grisak}
\orcid{0009-0005-2498-837X}
\affiliation{%
  \institution{Colorado School of Mines, Department of Mechanical Engineering}
  \city{Golden}
  \state{CO}
  \postcode{80401}
  \country{USA}}
\email{agrisak@mines.edu}

\author{Gabe Fierro}
\orcid{0000-0002-2081-4525}
\affiliation{%
  \institution{Colorado School of Mines, Department of Computer Science}
  \city{Golden}
  \state{CO}
  \postcode{80401}
  \country{USA}}
\email{gtfierro@mines.edu}

\renewcommand{\shortauthors}{C. Estelle Smith, \textit{et al.}}

\begin{abstract}
Because of the rapid development and increasing public availability of Generative Artificial Intelligence (GenAI) models and tools, educational institutions and educators must immediately reckon with the impact of students using GenAI.
Yet there is limited prior research on computing students' use and perceptions of GenAI. 
We surveyed all computer science majors in a small engineering-focused R1 university in order to: 
(1) capture a baseline assessment of how GenAI has been immediately adopted by aspiring computer scientists; 
(2) describe computing students' GenAI-related needs and concerns for their education and careers; and
(3) provide recommendations for GenAI policy development and educational tooling.
We present an exploratory qualitative analysis of this data and discuss the impact of our findings on the emerging conversation around GenAI and education.
\end{abstract}



\newcommand{\participanttable}{
\begin{table*}[]
\footnotesize

\begin{tabular}{|l|c|c|c|c|c|c|}
\hline
& \multicolumn{3}{c|}{\textbf{Undergraduate Students}} 
& \multicolumn{3}{c|}{\textbf{Graduate Students}} \\ 
\hline

{\textbf{Total Respondents}} & \multicolumn{3}{c|}{Total: $N=116$} & MS: 12 & PhD: 5 & Total: $N=17$ \\ 
\cline{2-7} 

& \multicolumn{1}{l|}{Years Enrolled} & \multicolumn{2}{l|}{$N$ (\% of Undergrad Sample)} & Years Enrolled & \multicolumn{2}{l|}{$N$ (\% of Grad Sample)} \\ \cline{2-7} 

& 0-1 & \multicolumn{2}{l|}{33 (28.4\%)} & 0-1 & \multicolumn{2}{l|}{3 (17.6\%)} \\ \cline{2-7} 

& 1-2 & \multicolumn{2}{l|}{43 (37.1)} & 1-2 & \multicolumn{2}{l|}{2 (11.8\%)} \\ \cline{2-7} 

& 2-3 & \multicolumn{2}{l|}{21 (18.1\%)} & 2-3 & \multicolumn{2}{l|}{1 (5.8\%)} \\ \cline{2-7} 

& 3-4 & \multicolumn{2}{l|}{18 (15.5\%)} & 3-4 & \multicolumn{2}{l|}{6 (35.3\%)} \\ \cline{2-7} 

& 4+ & \multicolumn{2}{l|}{1 (0.8\%)} & 4+ & \multicolumn{2}{l|}{5 (29.4\%)} \\ 
\hline

{\textbf{Frequency of Use}} & LLM & Code Generator & Image Generator & LLM & Code Generator & Image Generator \\ 
\cline{2-7} 

\rule{0pt}{0.30in} 
\begin{tabular}[c]{@{}l@{}} 
    \textit{X-axis from left to right:} \\
    only for fun or curiosity; \\ 
    never; once or twice ever; \\ 
    regularly (once or twice/week); \\ 
    nearly everyday 
\end{tabular}

& \raisebox{-0.18in}{\includegraphics[width=0.7in,height=0.4in]{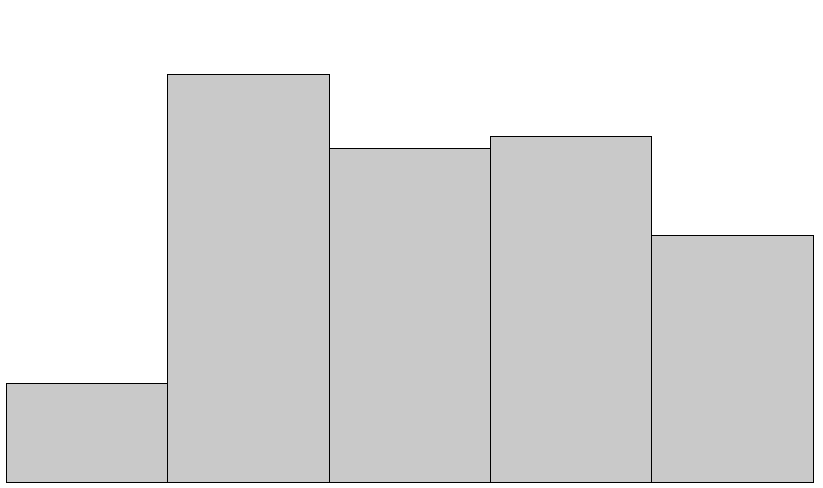}} 
& \raisebox{-0.18in}{\includegraphics[width=0.7in,height=0.4in]{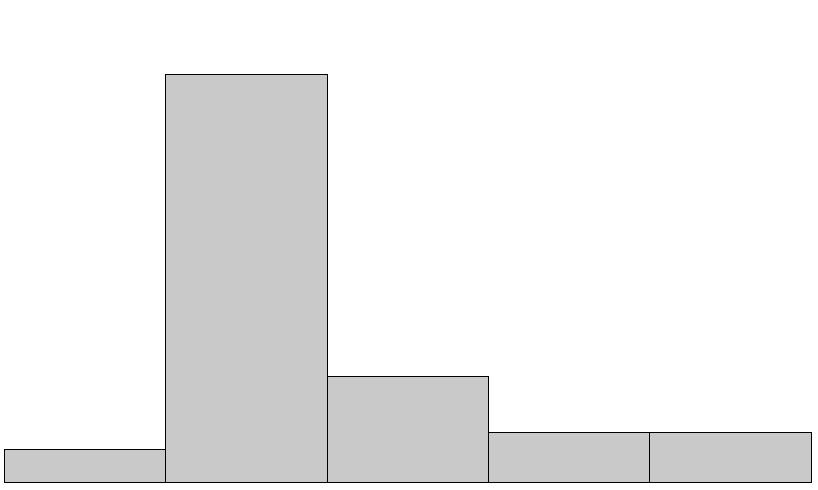}}  
& \raisebox{-0.18in}{\includegraphics[width=0.7in,height=0.4in]{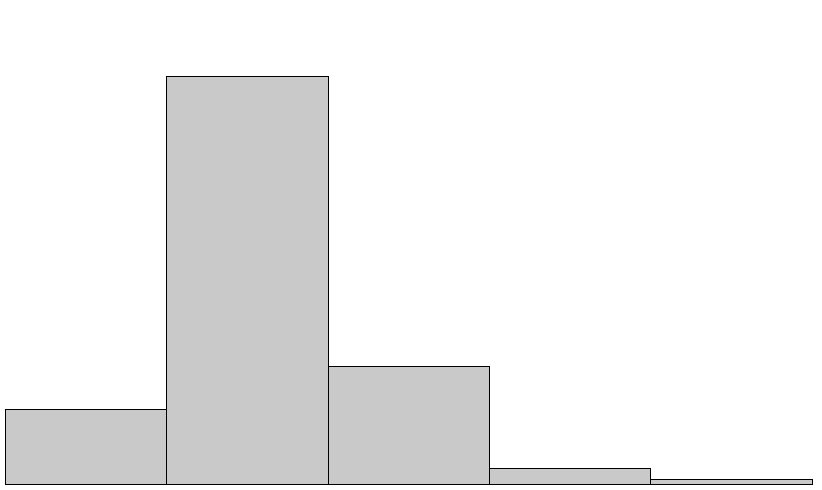}}  
& \raisebox{-0.18in}{\includegraphics[width=0.7in,height=0.4in]{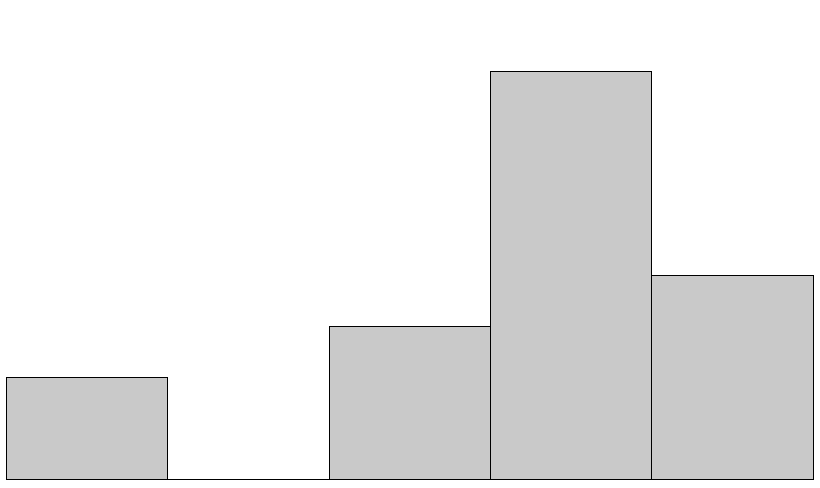}} 
& \raisebox{-0.18in}{\includegraphics[width=0.7in,height=0.4in]{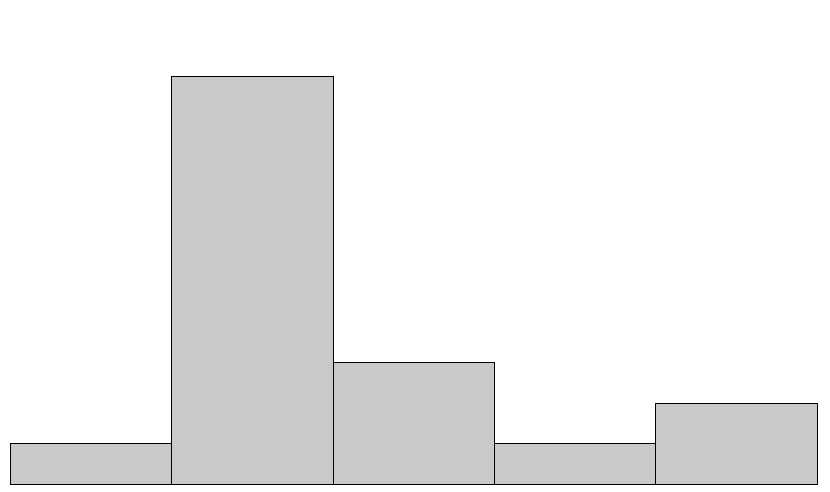}} 
& \raisebox{-0.18in}{\includegraphics[width=0.7in,height=0.4in]{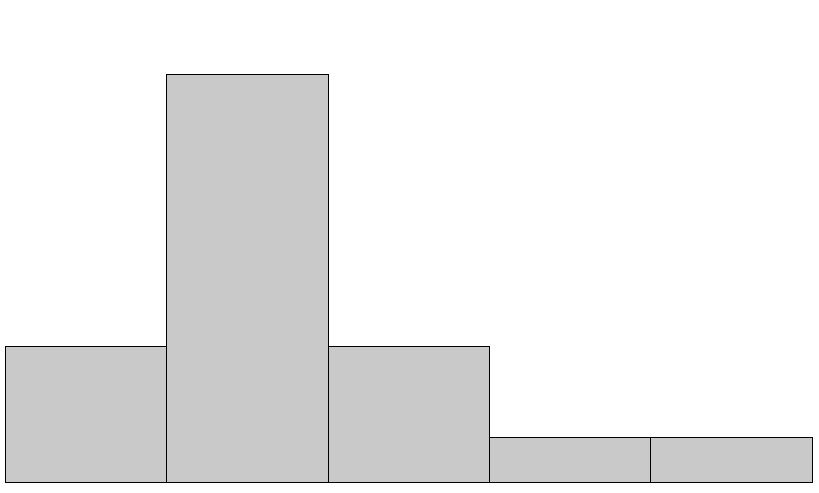}} \\
\hline

\rule{0pt}{0.30in} 
\begin{tabular}[c]{@{}l@{}} 
    \textbf{Perceived Benefit to CS} \\ 
    1 (left): extremely damaging \\ 
    10 (right):extremely beneficial \\ 
    \phantom{addrow} 
\end{tabular}
& \multicolumn{3}{c|}{\raisebox{-0.15in}{
    \includegraphics[width=1.5in,height=0.4in]{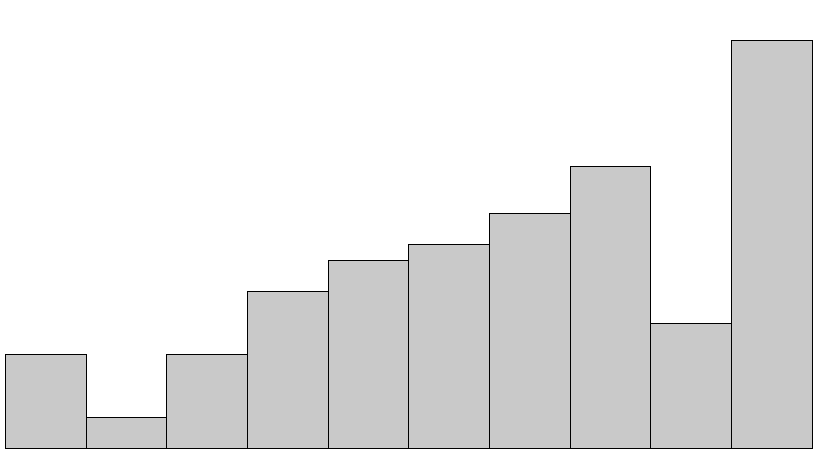}
}} 
& \multicolumn{3}{c|}{\raisebox{-0.15in}{
    \includegraphics[width=1.5in,height=0.4in]{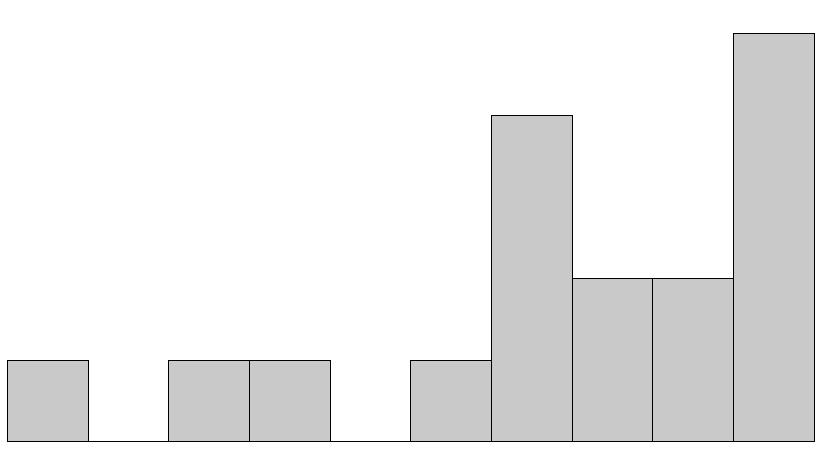}
}} \\ 
\hline

\end{tabular}

\caption{Summary of participant demographics and key quantitative survey questions. Histograms are included for visual synthesis of students' frequency of use of GenAI and their ratings of how beneficial GenAI will be to the field of computer science; the results section reports numbers of participants in each category.}

\label{tab:participants}

\end{table*}

}

\newcommand{\Qonecodes}{

\begin{figure}[tb]
    \centering
    \includegraphics[width=0.45\textwidth]{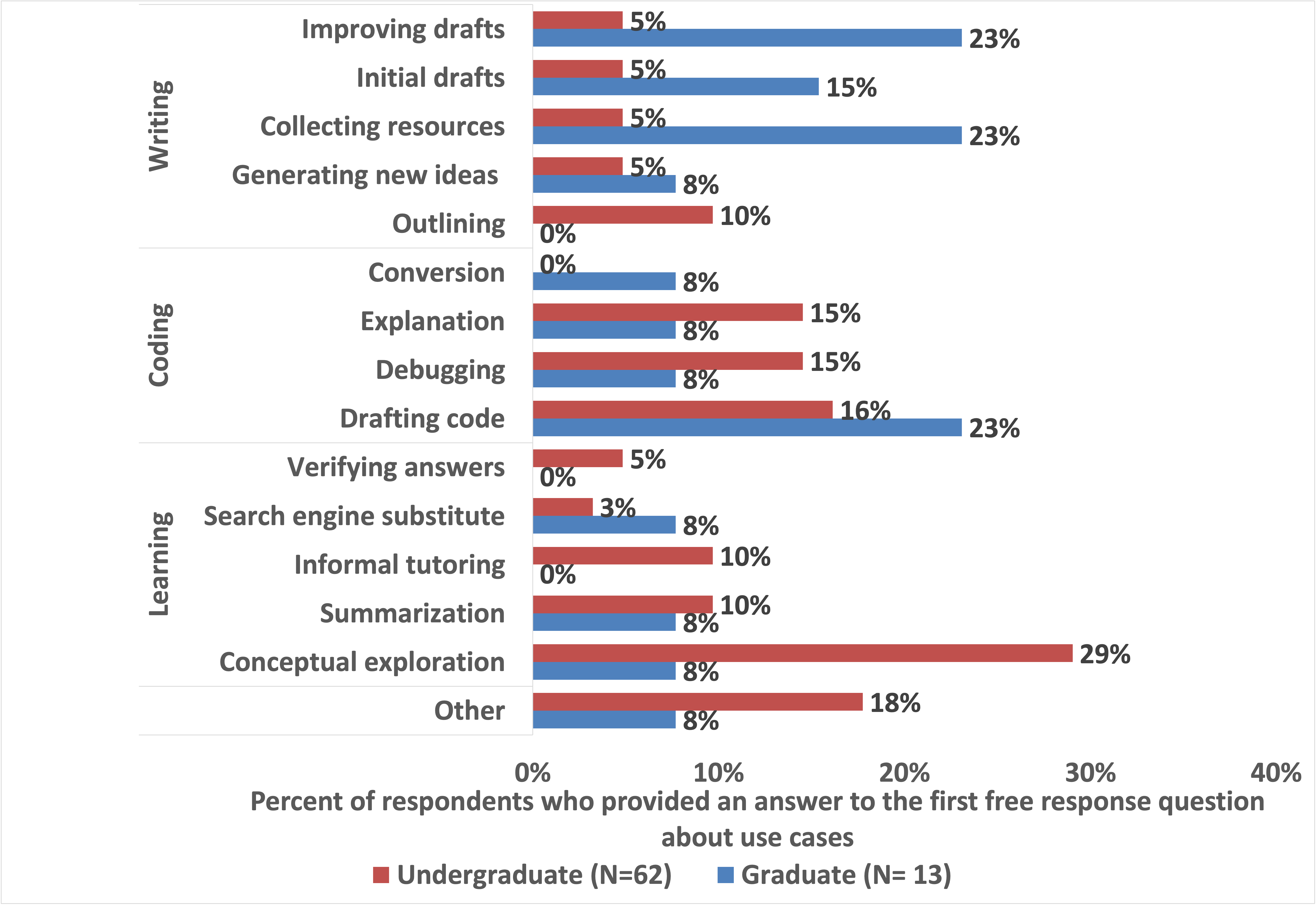}
    \caption{Summary of RQ1 codes applied.}
    \label{fig:Q1Codes}
\end{figure}

}

\newcommand{\Qtwothreecodes}{

\begin{figure}[tb]
    \centering
    \includegraphics[width=0.45\textwidth]{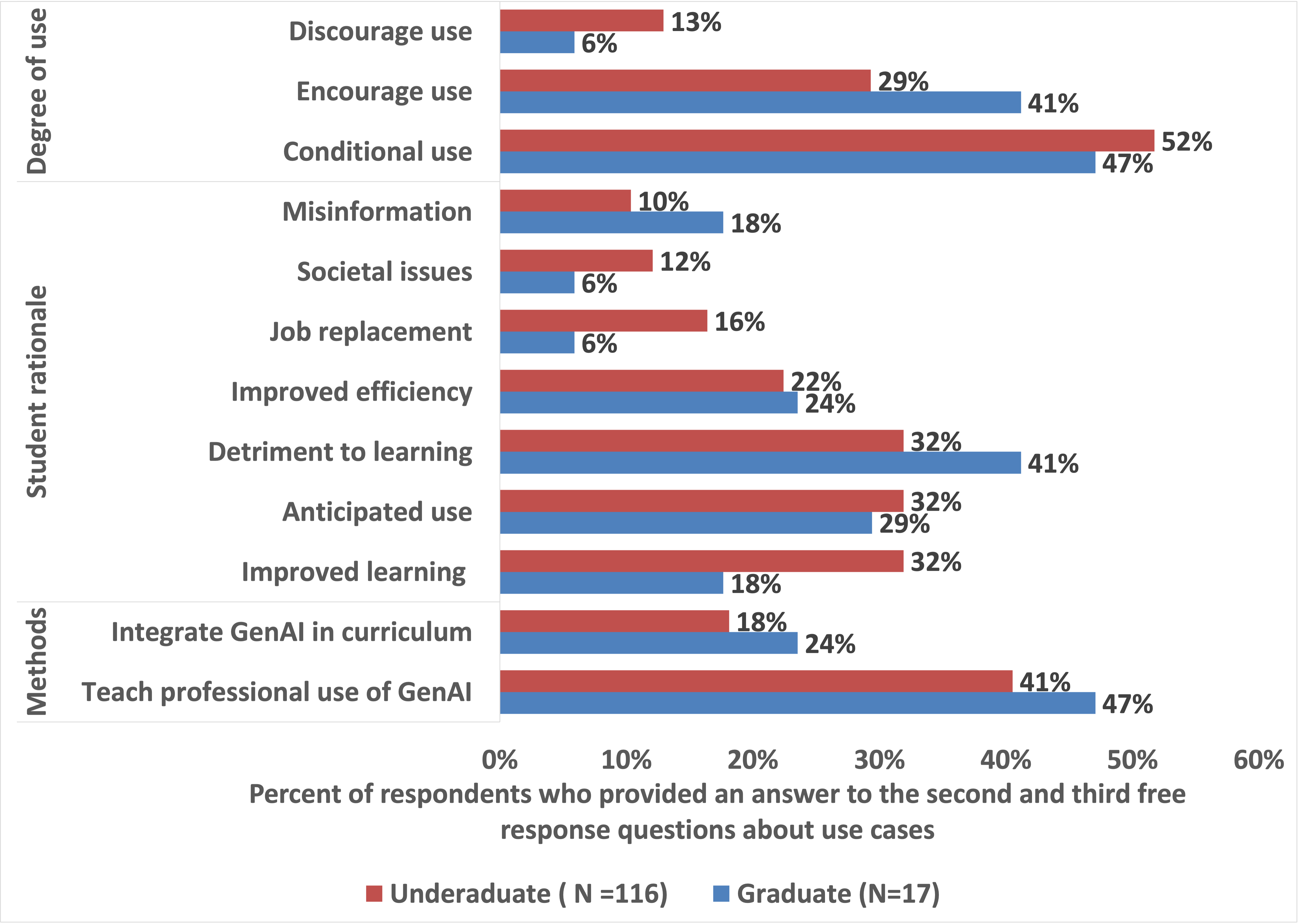}
    \caption{Summary of RQ2 codes applied.}
    \label{fig:Q23Codes}
\end{figure}

}

\newcommand{\IRRscores}{
\begin{table}[]
\footnotesize
\begin{tabular}{|lclcl|}
\hline
\multicolumn{4}{|c|}{\textbf{RQ IRR Scores}}                                                                                                                                             \\ \hline
\multicolumn{2}{|p{1cm}|}{RQ1}                                                                  & \multicolumn{2}{p{1cm}|}{RQ2}                                                           \\ \hline \hline
\multicolumn{1}{|c|}{Code}                                   & \multicolumn{1}{c|}{Score}  & \multicolumn{1}{c|}{Code}                                  & Score                 \\ \hline
\multicolumn{1}{|l|}{Generating new ideas} & \multicolumn{1}{c|}{0.880}  & \multicolumn{1}{l|}{Discourage use}                        & 0.780                 \\ \hline
\multicolumn{1}{|l|}{Outlining}                              & \multicolumn{1}{c|}{0.897}  & \multicolumn{1}{l|}{Conditional use}                       & 0.627                 \\ \hline
\multicolumn{1}{|l|}{Collecting resources}                   & \multicolumn{1}{c|}{0.636}  & \multicolumn{1}{l|}{Encourage use}                         & 0.629                 \\ \hline
\multicolumn{1}{|l|}{Initial drafts}                         & \multicolumn{1}{c|}{0.772}  & \multicolumn{1}{l|}{Improved efficiency}                   & 0.622                 \\ \hline
\multicolumn{1}{|l|}{Improving drafts}                       & \multicolumn{1}{c|}{0.948}  & \multicolumn{1}{l|}{Improved learning outcomes}            & 0.618                 \\ \hline
\multicolumn{1}{|l|}{Drafting code}                          & \multicolumn{1}{c|}{0.870}  & \multicolumn{1}{l|}{Anticipated use}                       & 0.671                 \\ \hline
\multicolumn{1}{|l|}{Debugging}                              & \multicolumn{1}{c|}{0.929}  & \multicolumn{1}{l|}{Misinformation}                        & 0.644                 \\ \hline
\multicolumn{1}{|l|}{Conversion}                             & \multicolumn{1}{c|}{1.000*} & \multicolumn{1}{l|}{Societal issues}                       & 0.645                 \\ \hline
\multicolumn{1}{|l|}{Explanation}                            & \multicolumn{1}{c|}{0.812}  & \multicolumn{1}{l|}{Discipline replacement}                & 0.636                 \\ \hline
\multicolumn{1}{|l|}{Conceptual exploration}                 & \multicolumn{1}{c|}{0.682}  & \multicolumn{1}{l|}{Detriment to learning/authenticity} & 0.616                 \\ \hline
\multicolumn{1}{|l|}{Summarization}                          & \multicolumn{1}{c|}{0.889}  & \multicolumn{1}{l|}{Teach professional use of GenAI}       & 0.630                 \\ \hline
\multicolumn{1}{|l|}{Informal tutoring}                      & \multicolumn{1}{c|}{0.749}  & \multicolumn{1}{l|}{Integrate GenAI in curriculum}         & 0.723                 \\ \hline
\multicolumn{1}{|l|}{Search engine sub.}             & \multicolumn{1}{c|}{0.636}  & \multicolumn{1}{l|}{}                                      & \multicolumn{1}{l|}{} \\ \hline
\multicolumn{1}{|l|}{Verifying answers}                      & \multicolumn{1}{c|}{0.796}  & \multicolumn{1}{l|}{}                                      & \multicolumn{1}{l|}{} \\ \hline
\multicolumn{1}{|l|}{Other}                                  & \multicolumn{1}{c|}{0.668}  & \multicolumn{1}{l|}{}                                      & \multicolumn{1}{l|}{} \\ \hline
\multicolumn{1}{|l|}{Personal}                               & \multicolumn{1}{c|}{1.000}  & \multicolumn{1}{l|}{}                                      & \multicolumn{1}{l|}{} \\ \hline
\multicolumn{1}{|l|}{Academic}                               & \multicolumn{1}{c|}{0.628}  & \multicolumn{1}{l|}{}                                      & \multicolumn{1}{l|}{} \\ \hline
\multicolumn{1}{|l|}{Mixed}                                  & \multicolumn{1}{c|}{0.627}  & \multicolumn{1}{l|}{}                                      & \multicolumn{1}{l|}{} \\ \hline
\multicolumn{1}{|l|}{Unknown}                                & \multicolumn{1}{c|}{0.671}  & \multicolumn{1}{l|}{}                                      & \multicolumn{1}{l|}{} \\ \hline
\multicolumn{4}{|c|}{* Indicates a category no coders selected}                                                                                                                 \\ \hline
\end{tabular}
\end{table}
}

\newcommand{\learning}{
\begin{table*}[]
\begin{tabular}{p{0.45\textwidth}|p{0.45\textwidth}} \textbf{Detriment to Learning \& Authenticity} & \textbf{Improved Learning Outcomes} \\ \hline \hline
\textit{``I believe that as these AIs become smarter and smarter, they will be very harmful to learning environments. I worry that AI will simply write it's own programs, and that software engineering will become completely obsolete in the next 3 decades, as the last line of human code written will be putting the finishing touches on an AI coder.''} & \textit{``AI is really powerful for learning, such as if you don't understand a math concept, you can ask it to explain how to do it, and ask for more and more details if needed. It is good for inspiration but shouldn't be used to talk for you. I've successfully used it many times to decipher an error message, explain functionality, and even tried to get it to write code blocks.''} \\  \hline
\textit{``It is too powerful of a tool and can be too much of a crutch that students can rely on. I have known students who have used it on every assessment, very blindly following whatever instructions it gave. They couldn't even justify why it was wrong or identify when it had made a flagrant error.''
} &  \textit{``Instructors should encourage the use of AI as a tool to enhance a student's education, not as a tool that does the student's task of learning for them. Students should use AI to help debug their code, learn from their mistakes, and learn new programming techniques and tools, not to generate all the code for them.'' 
} \\ \hline

\textit{``It will limit the growth and knowledge someone would be able to achieve without the same capabilities.''}
& \textit{``It serves as a valuable learning tool that helps students understand complex concepts, generate ideas, and receive feedback.''} \\ \hline
\end{tabular}
\caption{RQ2 examples of student data demonstrating tensions between the potential for GenAI to damage or improve learning.}
\label{tab:learning}
\end{table*}
}



\maketitle

\section{Introduction}\label{sec:intro}

Given the increasing public availability of Generative Artificial Intelligence (GenAI), today's computing students now have immediate access to a new class of tools that stand to transform their learning outcomes and career prospects. 
Existing educational tools have been built and studied for targeted purposes such as tutoring~\cite{anderson_intelligent_1985}, visualizing~\cite{robbins_gilp_2023}, or explaining code~\cite{perez_rediscovering_2020}. 
Yet students are also usually provided with instructional training and access to such tools, since their use may be required for particular learning outcomes. 
On the other hand, GenAI is \textit{general purpose} and can generate many different types of content (\textit{e.g.,} text, code, images,  music, speech, music) using only natural language prompts. 
GenAI offers a lower barrier to entry than existing fit-for-purpose tools and can be accessed by students without the instructor as an interlocutor.
This informs our guiding question: \textit{how might this rapidly evolving state of affairs w.r.t. GenAI reframe prior questions on tooling in computing education, and how might educational policies best support the changing educational and career needs of students?}


Since the release of OpenAI's ChatGPT in November 2022, GenAI-based technologies have rapidly entered the public consciousness, with extensive impacts across education and industry~\cite{khan_harnessing_2023,lim_generative_2023}.
In CS education, concerns have been raised for how the code-generating capacity of GenAI might interfere with the learning process at the heart of many CS classrooms~\cite{denny_computing_2023}. 
To ensure that pedagogical uses of AI are beneficial, the US Department of Education recommends that 
stakeholders collaborate to design AI in alignment with modern learning principles~\cite{cardona_artificial_2023}; many curricula are actively integrating AI literacy~\cite{casal-otero_ai_2023,celik_towards_2023}. 
Institutions are now also releasing guidelines specifically for GenAI, with some banning or allowing GenAI indiscriminately, and others leaving it to instructors' discretion~\cite{caulfield_university_2023}.

As key stakeholders impacted by such policies, students' needs and concerns should be central to the development of GenAI policies and tools. 
However, at this seminal moment in the public adoption of GenAI, no systematic research has yet captured students' usages of and perspectives on the role of GenAI in computing education.
Therefore, this study poses two research questions:

\begin{itemize}
    \item \textbf{RQ1:} With limited guidelines, guardrails, or pre-planning, how did computing students adopt GenAI-based tools during the Spring 2023 academic semester? 
    \item \textbf{RQ2:} How do computing students envision the role of GenAI within their education and their future careers?
\end{itemize}

To address these questions, we surveyed computer science majors at a small engineering-focused R-1 university in the USA.
We found that most students have tried GenAI tools (\textit{esp.} LLMs) for a variety of writing, coding, and learning use cases.
Moreover, students tend to view GenAI tools as beneficial to the field of computing.
In our discussion, we synthesize these results to discuss how educators can optimize GenAI-based policies and tools for the educational and professional needs of students. 




\section{Related Work}
\subsection{Existing Tooling in CS Education}
CS education has long grappled with questions around the nature and role of automation and tools in education---\textit{e.g.,} Online Python Tutor~\cite{guo_online_2013}, plug-ins to assess student-IDE interactions~\cite{lyulina_tasktracker-tool_2021}, interactive E-books~\cite{risha_stepwise_2021} or algorithm visualizations~\cite{robbins_gilp_2023}, web-based AI/ML literacy tools~\cite{chao_teach_2023}. Despite rising numbers of research publications on AI literacy since 2018~\cite{tenorio_artificial_2023}, there is limited research on GenAI specifically, since such tools became publicly available only in 2022. GenAI immediately evokes comparisons to work on intelligent tutoring systems (ITS)~\cite{anderson_intelligent_1985} and computer-aided instruction, which incorporate domain knowledge to provide automated customized feedback to students.
ITS research has focused both on data-driven improvement of feedback ~\cite{price_evaluation_2017} as well as investigating the nature of the interaction between student and ``tutor''~\cite{merrill_effective_1992,anderson_cognitive_1995}.
Conversational agents are another type of tool that employ natural language dialogues, a form of communication that is more intuitive for most students~\cite{perez_rediscovering_2020}.
These agents can provide instant and informative responses~\cite{song_interacting_2017}, improve student comprehension~\cite{okonkwo_python-bot_2020}, and are capable of providing personalized assistance that can be difficult for human instructors~\cite{gonda_chatbot_2019}.
Tutoring systems that use conversational agents for STEM education have also been found to yield learning gains comparable to trained human tutors~\cite{graesser_electronixtutor_2018}.
Bayesian Knowledge Tracing (BKT) can be also used in ITS to model each learner's mastery and improve predictions of student success~\cite{hutchison_individualized_2013}.

Assistive tools can improve student experience, but they might also hinder effective learning.
One routine concern for CS educators is cheating by using websites like Chegg.com or StackOverflow.com to copy solutions without properly learning the material~\cite{manoharan_contract_2020}.
This form of ``contract cheating'' is difficult to detect because students can purchase and find custom solutions to their assignments~\cite{manoharan_contract_2020}.
The widespread availability and adoption of GenAI systems drastically re-frames many of these prior questions and concerns.
Not only are GenAI systems capable of synthesis across a wider range of tasks and domains, but their speed of adoption has given educational institutions little time to respond in a thoughtful manner.
These systems also make mistakes and can confidently present incorrect information in a manner that previous tools do not---consequently transmitting false beliefs to human users~\cite{kidd_how_2023}.

\subsection{The Emergence of Generative AI}
We consider three broad categories of GenAI in this work: 

\begin{itemize}
    \item Large Language Model (LLM) chatbots (\textit{e.g.,} ChatGPT, Bard, Bing Chat) in standalone conversational user interfaces.
    \item LLM Code Generators (\textit{e.g.,} GitHub Co-Pilot), which are code generation \& auto completion tools integrated within code development environments.
    \item Image Generators (\textit{e.g.,} Dall-E, Midjourney)
\end{itemize}

GenAI models can generate natural language text that imitates human text with high levels of coherence, complexity, and diversity~\cite{openai_gpt-4_2023}.
The novelty of GenAI tools and the diversity of tasks they can perform, combined with the unexplainability of AI~\cite{sun_investigating_2022}, has created difficulty in informing educators about how to interact with GenAI tools. 
Although some instructors have used them to assist with tasks like lesson planning or creating rubrics~\cite{cooper_examining_2023}, there is a general lack of understanding about the functionality, limitations, and usage of these tools.
As a result, educators are struggling to catch up with students who are exploring GenAI on their own~\cite{seo_impact_2021, zhai_review_2021}.
Some institutions are already adopting custom GenAI tools in computing classes~\cite{pezzone_harvard_2023}.
There is an urgent need for research into understanding the role of GenAI in computing education to encourage positive learning outcomes.

\participanttable

\subsection{Student-Centered Policy Development}
Institutions must develop policies to address AI concerns~\cite{conner_accountability_2011}.
The US Department of Education issued a 2023 report emphasizing the need for designing AI interventions based on modern learning principles, strengthening trust, involving educators, appropriately addressing contextual considerations, and developing effective guidelines and guardrails~\cite{cardona_artificial_2023}. 
Given the potential benefits and risks, guidance on the responsible use of GenAI in particular is now needed in this transitional time~\cite{lau_ban_2023, chan_comprehensive_2023}.
Some US institutions have already released guidelines on using GenAI in the classroom~\cite{caulfield_university_2023}.
However, it is unknown whether or how student perspectives have been considered during their development. 
We position the perspectives, needs, and concerns of today's computing students as integral to the formation of GenAI-related policies and tooling environments because of their role in the efficacy, alignment, and facilitation of these policies.
At the time of the survey, our institution had not yet released institution-wide policies.
To help inform policy development, this study therefore systematically captured and assessed computing students' emergent usages and perceptions of GenAI at the end of the Spring 2023 semester,
amidst the initial wave of hype and adoption of GenAI.



\section{Methods}\label{sec:methods}
We surveyed all CS majors in a small USA-based engineering-focused R1 university. Whereas some universities do not count majors until the second year, students can arrive in the CS major at this university, so responses are from freshman undergraduates up through advanced PhDs. This section describes our survey recruitment, design, and analysis. Our study was reviewed by our institution's ethics review board and exempted from IRB review.

\subsection{Survey Recruitment}
We emailed the survey to list-servs for all undergraduates and graduates.
Beyond standard survey limitations of opt-in bias or possibly inaccurate self-assessments, the primary limitation of this study is that students may have withheld or misrepresented information about behaviors perceived as cheating for fear of repercussions.
To counteract this, we used messaging encouraging honesty and assured students that responses would not be tied to their identity.
We collected no identifying information and only two demographic details (undergraduate, masters, or doctoral student status; \# years enrolled).
At the end, they could opt-in to complete a separate form and enter their email in a drawing for one of four \$25 gift cards.


\subsection{Survey Design}
The survey first presented consent and eligibility information. 
We then asked students to rate their frequency of use of LLMs, code generators, and image generators for classes or professional efforts.
Since there were no institution-wide policies, we also asked whether their individual classes had GenAI policies in Sp23, and for student TAs, if they had come across suspected AI-generated student submissions.
We asked students to rate whether they think Gen-AI will be extremely harmful (1) to extremely beneficial (10) to the field of computing.
Three free response questions inquired: (a) how students have used GenAI; (b) how they feel it should be used in education; and (c) what their concerns are for future workplaces.
\S\ref{sec:results} contains the verbatim free response question text.

\subsection{Survey Analysis}\label{sec:DCA}
Data were exported from QuestionPro software into a CSV file.
We used standard \texttt{pandas} and \texttt{scikit-learn} Python packages to compute descriptive statistics and perform the statistical tests listed in results. 
To analyze free response questions, we utilized directed content analysis~\cite{hsieh_three_2005}.
We inductively developed detailed codebooks to capture emergent themes. 
Across six rounds of iteration, six human coders manually coded subsets of the data, continuously discussing disagreements and refining code definitions until consensus was achieved. 
We used Inter-Rater Reliability (IRR) scores (Krippendorff's alpha) to guide refinements at each round until all IRR scores were greater than 0.6 (a threshold that establishes good agreement across coders). 
Finally, individual coders coded all responses according to the finalized codebooks. 
\S\ref{sec:results} includes abbreviated descriptions of codes; the complete codebooks include full code definitions along with data examples. 
For replicability, all codebooks are available as a supplemental material at \underline{\textcolor{blue}{\href{https://bit.ly/SIGCSE-GenAI-codebooks}{bit.ly/SIGCSE-GenAI-codebooks}}}. Throughout results, \textbf{bold typesetting} indicates a high-level codebook category; \textit{italic typesetting} indicates an individual code.

\subsection{Sample Description}
Table~\ref{tab:participants} summarizes participants and two key quantitative questions. We received 133 responses from eligible consenting students. Of these, $N=116$ (87.2\%) are undergraduate students; this equates to 12\% of the total undergraduate population in the department at the time of the survey. $N=17$ (12.8\%) are graduate students; this equates to 7.6\% of the graduate population at the time of the survey. We acknowledge the limitation that this sample is not representative. Therefore we position our results primarily as qualitative and exploratory, and the distributions of codes applied may or may not be representative of the true distributions. Nonetheless, these codes accurately describe students' emergent use cases and perspectives. Future work should build on these results with larger samples across different types and sizes of institutions.
\Qonecodes

\section{Results}\label{sec:results}
\subsection{RQ1: Students' adoption of GenAI in Sp23}

Students reported using LLM chatbots more frequently than code or image generators. For instance, $N=24$ (18.0\%) of students use LLMs everyday; $N=36$ (27.1\%) regularly (once or twice per week); $N=30$ (22.6\%) once or twice ever; $N=33$ (24.8\%) never; or $N=10$ (7.5\%) only for fun or curiosity. Fewer have used code generators; $N=11$ (8.3\%) of students use code generators everyday; $N=10$ (7.5\%) regularly; $N=22$ (16.5\%) once or twice ever; $N=83$ (62.4\%) never; or $N=7$ (5.3\%) only for fun. Interestingly, only 36.1\% have ever reported trying an image generator, thus we omit further discussion of image generators.

At our institution, there were no institution-wide guidelines during Spring 2023. 12.8\% of participants reported that at least one class had a formal GenAI policy in the syllabus; 23.3\% that instructors stated a formal policy \textit{not} in the syllabus; 30.1\% that instructors mentioned only loose guidance; and 33.8\% that there was no discussion of GenAI. Of $N=31$ students who indicated that they were TAs, $N=13$ (41.9\%) did not believe they had encountered AI-completed assignments; $N=11$ (35.5\%) were unsure; and $N=7$ (22.6\%) believe or know they received AI-generated assignments. 

In order to understand how computing students used GenAI, we asked the following optional question: \emph{If you have ever used any GenAI-based tool(s) for your classes, research, and/or professional efforts, please tell us about how you have used them, why you used them, and how you feel about your use of these tool(s).}
75 students (56.4\% of the full sample) submitted an answer.
Directed content analysis (~\S\ref{sec:DCA}) yielded Tables 1-4 in the supplemental \underline{\textcolor{blue}{\href{https://bit.ly/SIGCSE-GenAI-codebooks}{codebooks}}} which structured this RQ1 analysis.
Figure~\ref{fig:Q1Codes} visualizes the distribution of codes applied according to normalized percentages of responses from undergrad \textit{v.s.} graduate students; the remainder of results reports counts of codes applied across all respondents.

No respondents reported using GenAI to fully complete assignments for them. Rather, they described how their use of GenAI tools supported three different categories of use cases for \textbf{writing}, \textbf{coding}, and \textbf{learning}. We labeled the \textbf{context} of these use cases, and found that the majority were either \textit{academic} ($N=29$) (\textit{i.e.,} for coursework or research) or \textit{unknown/not-specified} ($N=31$). Although no participants described purely \textit{professional} use cases, some reported \textit{mixed} ($N=7$) contexts spanning academic/professional. $N=1$ respondent described a \textit{personal} use context.

\textbf{Writing:} 21 respondents used GenAI tools for writing support. $N=4$ students used GenAI to \textit{generate new ideas for consideration} to help them overcome writer's block, diversify their thinking, or figure out what to write about. $N=6$ mentioned \textit{collecting resources} to support their writing---\textit{e.g.,} \textit{``If I formed a new point or thought that I didn't find a quote for on my first round annotating, ChatGPT would be very useful to help me find useful evidence or decide on a quote to integrate into my paper.''} They also described use cases to help with the mechanics of writing, including \textit{outlining} ($N=6$), \textit{generating initial drafts} ($N=5$), or \textit{improving drafts} ($N=6$) of papers and emails. For example, \textit{``Sending an email to my professor vs. sending an email to a friend will be different in terms of format and word choice, and Quillbot helps in this kind of scenario.''}\footnote{Quillbot is a writing assistant tool that is built on LLMs.}

\Qtwothreecodes

\learning

\textbf{Coding:} 23 respondents reported using GenAI to help them understand, create, or fix code. $N=13$ respondents reported \textit{drafting code} that they could then verify, modify, and complete. Examples include: \textit{``I use ChatGPT and Copilot regularly to help write code quicker for many more mundane implementation tasks.''}; \textit{``I view GPT as a calculator for coding.''} Another common use case was \textit{explanation} ($N=10$), \textit{i.e.,} providing code snippets to an LLM for an explanation of code behavior. For example, \textit{``I used Co-Pilot on a bit of code I was stuck on, and then I used ChatGPT to explain why Co-Pilot did what it did.''} Students also used GenAI for \textit{debugging} ($N=10$) to find and fix bugs in their own code. \textit{``I use ChatGPT to help me debug my code. it's quicker than crawling through stackoverflow forums and it gives a very detailed explanation of why my code is wrong and how the new way is better so i feel like i’m learning.''} $N=1$ respondent also used GenAI for \textit{conversion} of code from one language to another.

\textbf{Learning:} One key result is that 32 respondents described uses of GenAI for self-described improvement of their learning. The code \textit{conceptual exploration} ($N=19$) was highly prominent: students used LLMs to ask about course topics and gain supplementary perspectives or alternative wordings and thought processes from those provided by instructors, thereby deepening their knowledge. Examples include: \textit{``Sometimes I ask it about a concept in class and it explains it to me and knows how to dumb it down for me.''}; \textit{``I used it to explain a topic that our professor didn't speak much about.''}; \textit{``[ChatGPT] helps to explain thought processes behind hard-to-understand concepts.''} Students also mentioned \textit{summarization} ($N=7$) of lengthy documents and \textit{informal tutoring} ($N=6$) as techniques that helped them digest and interact with course materials, esp. when instructors were unavailable. For example, \textit{``Sometimes [it is] hard to contact professors to get help with homework so it's nice to have something to help.''} Some students also used GenAI as a \textit{search engine substitute} ($N=3$) or as tool for \textit{verifying answers} ($N=3$) rather than generating them.

\subsection{RQ2: The role of GenAI in CS Education}

We asked students to rate how beneficial they feel GenAI will be to computer science. Table~\ref{tab:participants} visualizes the distribution of ratings. We observe that the distributions tend towards more positive evaluations: the average undergraduate rating is 6.78 ($SD=2.62$) and grad rating is 7.41 ($SD=2.67$).

We asked two additional required free response questions: \emph{What do you think the role of generative AI should be in higher education? For example, should instructors be trying to encourage or prohibit use of GenAI for students' coursework? How and why, or why not?}; and
\emph{What GenAI-related concerns do you have with regard to the workplace you will soon be entering, and how do you want instructors to prepare you for this workplace?}
We concatenated answers to these two questions and applied codes across both since there was substantial overlap in the content of these responses. All 133 participant answers were included in this analysis. 
Tables 5-7 in the supplemental \underline{\textcolor{blue}{\href{https://bit.ly/SIGCSE-GenAI-codebooks}{codebooks}}} structured this RQ2 analysis.
Figure~\ref{fig:Q23Codes} visualizes the distribution of codes applied. We coded three categories related to students' desired \textbf{degree of use} of GenAI in education, their \textbf{rationale} behind these opinions, and what \textbf{methods of implementation} could serve their educational needs.

\textbf{Degree of use:} 126 respondents reported opinions on what degree of GenAI usage would be most appropriate. Most called for \textit{conditional use} ($N=69$), meaning instructors should specify under what circumstances GenAI use is allowed, appropriate, and ethical.
Others wanted to \textit{encourage use} ($N=41$) without restrictions, while only a few wanted instructors to entirely \textit{discourage use} ($N=16$).

\textbf{Student rationale:} 111 responses also included rationales for their opinions. Importantly, there are split opinions on how GenAI may impact learning (see Table~\ref{tab:learning} for data examples): $N=44$ students were concerned about GenAI's potential \textit{detriment to learning and authenticity}, whereas $N=40$ felt GenAI could \textit{improve learning outcomes} through the types of use cases described in RQ1. Many students \textit{anticipated use} ($N=42$) of GenAI tools in their future careers and felt it would be necessary for them to learn them in order to be competitive and effective at their jobs. Yet students also voiced concerns about \textit{misinformation} ($N=15$) produced by GenAI, \textit{societal issues} ($N=15$) such as unethical use, intellectual property violations or plagiarism, privacy breaches, and unfair advantages or equitable access issues, or \textit{job replacement} ($N=20$).

\textbf{Methods of implementation:} 70 respondents mentioned methods such as \textit{teaching professional use of GenAI} ($N=55$) to prepare students with the specific GenAI-related skills that they will need for their future workplaces and/or \textit{integrating GenAI in the curriculum} ($N=30$) to cultivate effective and appropriate uses of GenAI to support learning during their education.

\paragraph{Exploratory Statistical Analysis} 
We queried whether any of the manually applied codes were interrelated using Pearson correlation coefficients; no coefficients exceeded 0.314, suggesting that the codes are not significantly correlated.
We used chi-squared tests of independence to examine relationships between qualitative codes and quantitative responses and report on interesting relationships with $p<0.01$. 
We found that students who used LLMs more frequently were more likely to rate the benefit of GenAI more highly ($p<.0008$) and to have RQ2 responses coded with \textit{encourage use} or \textit{conditional use} ($p<0.002$). 
Students who have not used GenAI provided lower ratings of its benefits ($p<.0001$) and were more likely to have RQ2 responses coded with \textit{discourage use} ($p<.0003$). 

\section{Discussion}
The widespread public availability of GenAI is driving a paradigm shift in computing education.
Our results demonstrate that many CS students rapidly adopted GenAI to support their \textbf{writing}, \textbf{coding}, and \textbf{learning}, even with most instructors providing little to no formal guidance on GenAI. 
Student quotes indicate that they view GenAI usage as \textit{``inevitable''} or \textit{``the new google''}; they tend to view GenAI as beneficial to CS, with many investing effort to use GenAI as a \textit{``parallel colleague.''}.
Our work has also surfaced a central tension.
Students who have used GenAI shared many positive use cases and experiences but are split as to what extent GenAI use should be guided or restricted, and to what extent it may improve or impair learning.
Educators and researchers alike should address these concerns directly; we offer some initial thoughts here.

\paragraph{Culture, policy, and tooling}
One first step is to understand and explore the emerging use cases in Fig.~\ref{fig:Q1Codes} and critically assess how such use cases relate to the learning objectives for different courses and developmental stage of students. 
Clarity from instructors and institutions is necessary to address internal and interpersonal tension within the student body on when and how to use GenAI---\textit{i.e.,} what constitutes cheating or academic dishonesty \textit{v.s.} an allowable and helpful use?
For example, if GenAI were used early in CS education, students might fully complete course assignments without developing a solid understanding of fundamental course concepts, hindering their ability to approach higher-level concepts later in their curriculum where GenAI might exceed its limit to help. 
Consequently, an intro CS instructor could hypothetically encourage uses such as \textit{concept exploration}, \textit{code explanation}, or \textit{answer verification} but expressly forbid \textit{drafting code} or \textit{debugging}, whereas a senior-level course might allow unrestricted use.
With GenAI tools now freely available online, creating a culture of honesty and accountability is essential to the success of such policies. 
Moreover, given that some institutions may develop custom in-house GenAI-based tooling~\cite{pezzone_harvard_2023}, future research and innovation should explore how to enforce guardrails within custom implementations that provide educators with the ability to technically prohibit certain use cases while allowing others. 
We believe that a harmonious balance in culture, policy, and tooling could ultimately improve educational outcomes, but many outstanding questions must be addressed in pursuit of this goal.

\subsection{Future Work}
\paragraph{Under what circumstances does GenAI improve v.s. harm learning?} Student perceptions of improved learning may or may not align with actual improved learning outcomes. Future research should explore how student outcomes (\textit{e.g.,} GPA, test scores, or other performance evaluations) are affected by different types of GenAI usage. 
Although our sample size is not representative, Fig.~\ref{fig:Q1Codes} is also suggestive that possible differences may exist in undergraduate \textit{v.s.} graduate use cases; understanding the role of experience \textit{w.r.t.} GenAI usage is another valuable direction for future work.
Moreover, what do students \textit{need} to learn, and how do we determine student success under this emerging GenAI paradigm?
Fact recall, memorization, and simple calculations may all be necessities for computing students, but these are aspects that AI can currently replace. 
Meanwhile, computing students need to be capable of critical thinking, logical reasoning, peer-to-peer relationships, communication, and other skills that are not easily replaced by GenAI. 
More broadly, the rise of GenAI usage is an opportunity to question the role of education and to reexamine long-held beliefs about student learning outcomes and assessment strategies.

\paragraph{How can GenAI be used safely, equitably, and sustainably?}
Student responses raised important concerns about misinformation, job replacement, and societal issues.
It is also unclear at the time of writing whether GenAI will become financially or environmentally sustainable longterm.
Future research must explore how to reduce misinformation presented to students and address accessibility, diversity, equity, and employment concerns.

\paragraph{How do CS instructors' needs and perspectives relate to students' perspectives on GenAI?} Our survey explored student perspectives, yet instructors are also vital stakeholders. Pressing challenges for the expansion of CS education include staffing issues, curricular capacity, school accountability pressure, and equitable access to CS coursework~\cite{bruno_four_2022}. Similar to students' concerns of job replacement, GenAI could potentially threaten the role of educators (\textit{e.g.,} tutors, lecturers, graders), yet it could also expand access to CS education. Future research should similarly capture instructors' perspectives, explore how to support them in utilizing GenAI effectively in their classrooms, and understand how to balance uses of GenAI to manage staffing challenges.

\section{Conclusion}
Our survey contributes a foundational snapshot of how students at an R1 USA-based CS department immediately adopted GenAI tools following their public release in 2022.
Results suggest a complex and evolving relationship with GenAI, in which students' emergent use cases are already impacting their learning processes and outcomes. 
We discuss how educators should design policies and tools that effectively use GenAI to benefit student learning while preparing them to utilize these resources in the workforce. 
Finally, we suggest future work that is needed to align CS curricula with the behaviors and educational needs of computing students in light of GenAI.

\section{Acknowledgements}
We thank all student participants who took this survey. We would also like to acknowledge James Frishkoff for creating a preliminary draft of the survey, James Sprow and Laila Barrett for help with formatting, Tracy Camp, Kathyrn Stolee, and Christine Liebe for feedback on a preliminary draft of the manuscript, and Iris Bahar for distributing the survey to the CS dept.

\bibliographystyle{ACM-Reference-Format}
\bibliography{generative_ai}

\appendix

\end{document}